\documentclass[aps,prl,twocolumn,showpacs,superscriptaddress,letterpaper,amsmath,amssymb]{revtex4} 

\usepackage{graphicx}  % needed for figures
\usepackage{dcolumn}   % needed for some tables
\usepackage{bm}        % for math
\usepackage{amssymb}   % for math
\usepackage{feynmf}%%%{feynmp}
\usepackage{slashed}
\def\gsim{\lower0.5ex\hbox{$\:\buildrel >\over\sim\:$}}
\def\lsim{\lower0.5ex\hbox{$\:\buildrel <\over\sim\:$}}

\newcommand{\be}{\begin{equation}}
\newcommand{\ee}{\end{equation}}
\newcommand{\bea}{\begin{eqnarray}}
\newcommand{\eea}{\end{eqnarray}}

\newcommand{\nbox}{{\,\lower0.9pt\vbox{\hrule \hbox{\vrule height 0.2 cm
\hskip 0.2 cm \vrule height 0.2 cm}\hrule}\,}}

\begin{document}

\thispagestyle{empty}
\vspace*{-3.5cm}

\vspace{0.5in}

%\begin{flushright}
%\today\\
%\end{flushright}
%\vspace{0.5in}
\title{Triangulating an exotic $T$ quark}

\begin{center}
\begin{abstract}
Limits on an exotic heavy quark $T$ are broadly generalized by considering the full range of
$T\rightarrow Wb, th$ or $tZ$ branching ratios. We combine
results of specific $T\rightarrow tZ$ and $T\rightarrow Wb$ searches
with limits on various
combinations of decay modes evaluated by re-interpreting other
searches. We find strong bounds across the entire space of branching
ratios, ranging from $m_T > 415$ GeV to $m_T > 557$ GeV at 95\% confidence level.
\end{abstract}
\end{center}

\author{Kanishka Rao}
\author{Daniel Whiteson}
\affiliation{UC Irvine, Irvine, CA }

\maketitle

\section{Introduction}

A fourth generation of fermions would be a natural extension to the
standard model of particle physics. Direct searches for a chiral
fourth-generation ($b'$ and $t'$), however, have yielded no evidence for a fourth
generation of fermions. Specifically, the CMS collaboration
has set limits  of $m_{t'} > 557$ GeV if BR($t'\rightarrow Wb$)=100\%~\cite{cmstpll}
and \mbox{$m_{b'}>611$ GeV~\cite{cmsbpll}} if BR($b'\rightarrow Wt$)=100\%.
Even if these branching ratios are reduced by off-diagonal mixing
terms, these analyses have complementary sensitivity which is nearly
impossible to escape if the fourth generation is chiral and decays  via
$W$-boson emission~\cite{flacco}. 

\begin{figure}[Ht]
\includegraphics[width=0.9\linewidth]{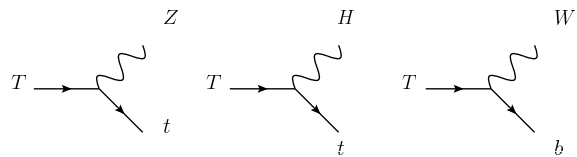}
\caption{Decay modes of a heavy exotic quark, $T$}
\label{fig:tp_diag}
\end{figure}

A fourth-generation quark, however, may be a vector particle ($T$)
which has exotic decays~\cite{tprime,jaas}, such as $T\rightarrow tZ$
or $th$, see Fig.~\ref{fig:tp_diag}.   Such a quark is a generic
feature~\cite{thiggs} of models in which the Higgs boson is a composite state, such
as models featuring a ``little
Higgs''~\cite{littlehiggs,littlehiggs2,littlehiggs3}.  The
contribution from the $T$ quark in such models is essential to cancel
the contributions to the Higgs mass from the top quark, keeping the
Higgs mass at the electroweak scale.

\begin{figure}[Ht]
\includegraphics[width=0.7\linewidth]{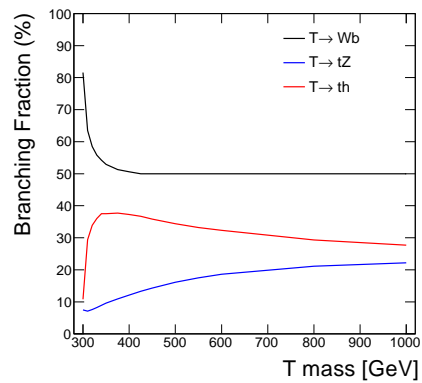}
\caption{ Branching ratio of $T$-quark decays to $Wb,tZ$ and $th$ vs $m_T$, from the model in Ref.~\cite{tprime}.}
\label{fig:br}
\end{figure}

The LHC collaborations have strong sensitivity to such a
quark~\cite{atlasstudy}. CMS performed a dedicated search
for this $T$ quark in data with 1 fb$^{-1}$, reporting
$m_T>475$ GeV~\cite{cmstz} if \mbox{BR($T\rightarrow tZ$)=100\%}, but no
 searches have been reported for the $T\rightarrow th$ mode, nor for
 models with realistic mixtures of decay modes, see Fig.~\ref{fig:br}.

In previous work, we reinterpreted an ATLAS search for $b'\rightarrow
tW$ which has  broad sensitivity for other heavy quark
modes; we set limits on the $T$-quark mass in the case of a realistic mixture
branching ratios, \mbox{$m_T>419$ GeV~\cite{basis}} if BR($T\rightarrow tZ$)=15\%,
BR($T\rightarrow th$)=35\%, BR($T\rightarrow Wb$)=50\%.

In this paper, we relax the assumptions which determine the branching
ratios as a function of mass  and explore the entire
 space of possible branching ratios, achieving an important
 generalization of the existing limits.

\section{ Direct Searches}

The CMS collaboration searched for a heavy chiral fourth-generation
$t'$ quark decaying via $Wb$ in the di-lepton decay mode using data
with 5.0 fb$^{-1}$ of luminosity.  We assume that this limit,
$m_T>557$ GeV at 95\% CL, is also
applicable to the exotic quark $T$ when it decays to $Wb$.
Other analyses have also been performed\cite{atlastplj,atlastpll,bat},
but we only consider the strongest limits.

In addition, CMS searched directly for $T$-quark production
specifically in the mode $T\rightarrow tZ$.  Assuming BR($tZ$) is
100\%, their analysis yields $m_T>475$ GeV at 95\% CL from data with 1
fb$^{-1}$ of integrated luminosity.

The ATLAS collaboration reported a search for heavy fourth-generation
down-type chiral quarks ($b'$) using data with 1
fb$^{-1}$~\cite{bprime} of integrated luminosity.   The
$b'$ decays via $tW$, leading to a final state with four $W$ bosons
and two $b$ quarks. The single-lepton mode is used, leaving three
hadronically decaying $W$ bosons. The ATLAS search makes use
of a novel technique for tagging boosted $W$ bosons by searching for
jet pairs with small angular separation. The analysis variables are the
jet multiplicity and $W$ boson multiplicity; it is a counting
experiment in nine bins: \mbox{$N_{jet}=(6,7,\ge 8) \times N_{W}=(0,1,\ge
2)$}. This analysis does not
directly set limits on $T$-quark models, but it is clear that if such particles were
produced, they would appear as an excess in this analysis.

Note that the precise aim of the original $b'\rightarrow tW$ analysis is not directly
relevant to the analysis presented in this paper, as we are reinterpretting it in another context.  However,
since the decay modes are similar ($b'\rightarrow tW$ vs $T\rightarrow
tZ$) in topology makes the reinterpretation more powerful and robust.

\section{Reinterpreting $b'\rightarrow tW$}

Given the clear sensitivity of the ATLAS $b'$ search to $T$-quark
production and decay, we reinterpret this analysis in the context of
$T$-quark models.

The ATLAS analysis uses a binned likelihood with nine bins in jet and
$W$-boson multiplicity: \mbox{$N_{jet}=(6,7,\ge 8) \times N_{W}=(0,1,\ge
2)$}. Until recently, reinterpretation of such a multi-bin analysis was
effectively impossible, as it would require publication by the
experiment of the complete likelihood details, including bin-to-bin
correlations. However, we showed in recent work~\cite{basis} that if
the template for a new signal model can be expressed as a linear combination of the
templates for models tested by the experiment, then limits on the new
model can be trivially derived.   This approach can be understood as
an interpolation strategy, and is valid when the dominant systematic
uncertainties are due to the background sources, or are similar
between the basis templates and the new signal model.

We exploit this approach to derive limits on $T$-quark production and
decay in a variety of decay modes.  We generate $T$ production and decay using {\sc madgraph}~\cite{madgraph},
use {\sc pythia}~\cite{pythia} to model showering and hadronization,
and use {\sc pgs}~\cite{pgs} to describe the detector response.  In every
case, we use $m_h=125$ GeV.

In the following sections, we find mixtures of the $b'\rightarrow tW$
templates which closely approximate templates for exotic $T$ quark
decays in $tZ, th, (tZ$ and $th), (Wb$ and $th),$ sor $(Wb, th,$ and $tZ)$
decay modes. In each case, we find the approximations to be imperfect
but reasonable.  Variations of the template mixtures which have
discrepancies of similar magnitude but opposite sign give similar results
to those we present, suggesting that the results are not highly
sensitive to these discrepancies. Uncertainties due to the imperfect
description are assessed.

\subsection{Decays to $tZ$}

If the $T$-quark decays exclusively to $tZ$, then the final state of
$t\bar{t}ZZ$ closely resembles the $ttWW$ final state of the $b'$
search, as the jet resolution does not allow us to distinguish between
$W\rightarrow qq'$  and $Z\rightarrow q\bar{q}$ decays, and the
mass window used ($70-100$ GeV)  encompasses most hadronic $Z$-boson decays.

\begin{figure}[Ht]
\includegraphics[width=0.44\linewidth]{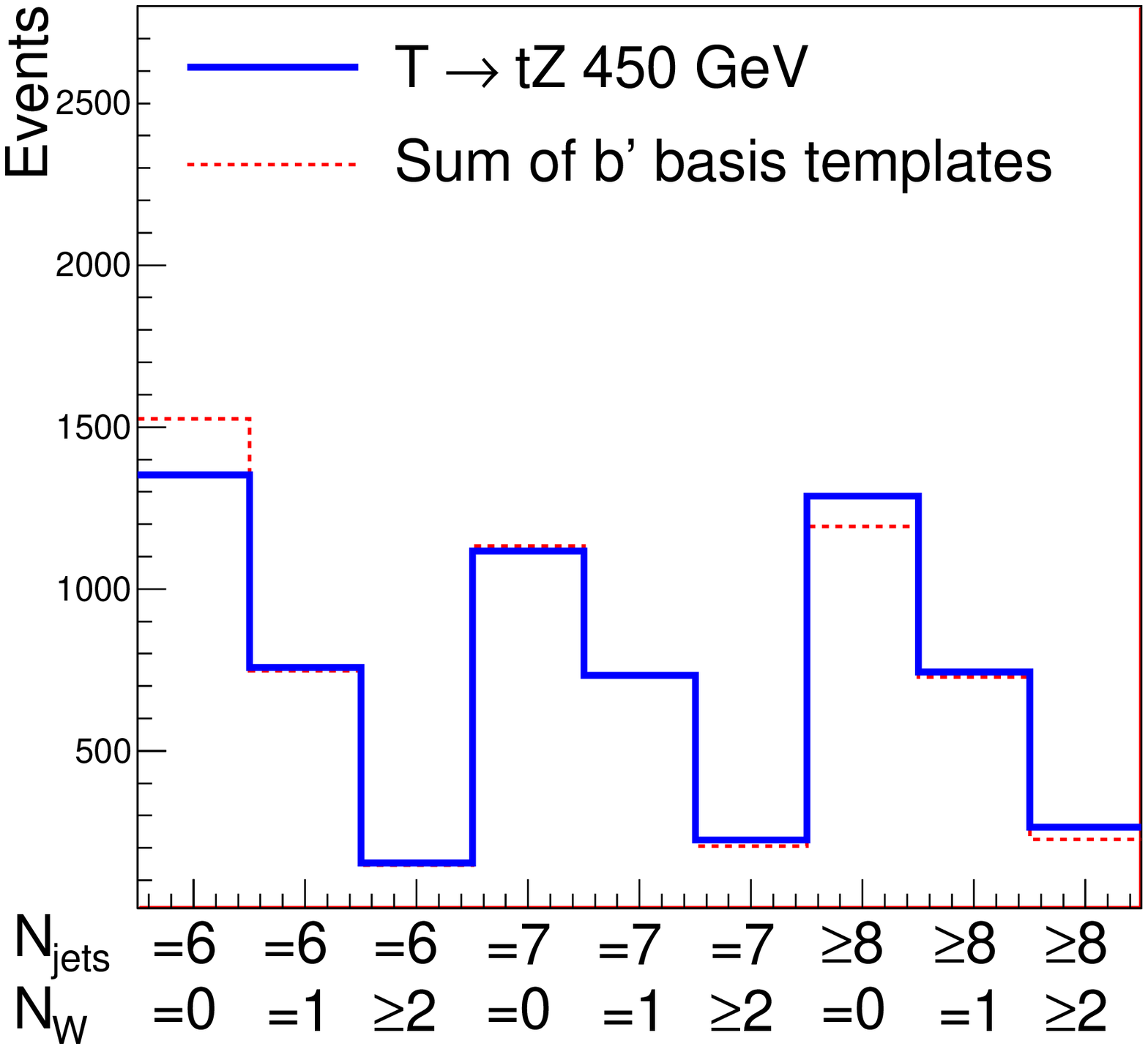}
\includegraphics[width=0.44\linewidth]{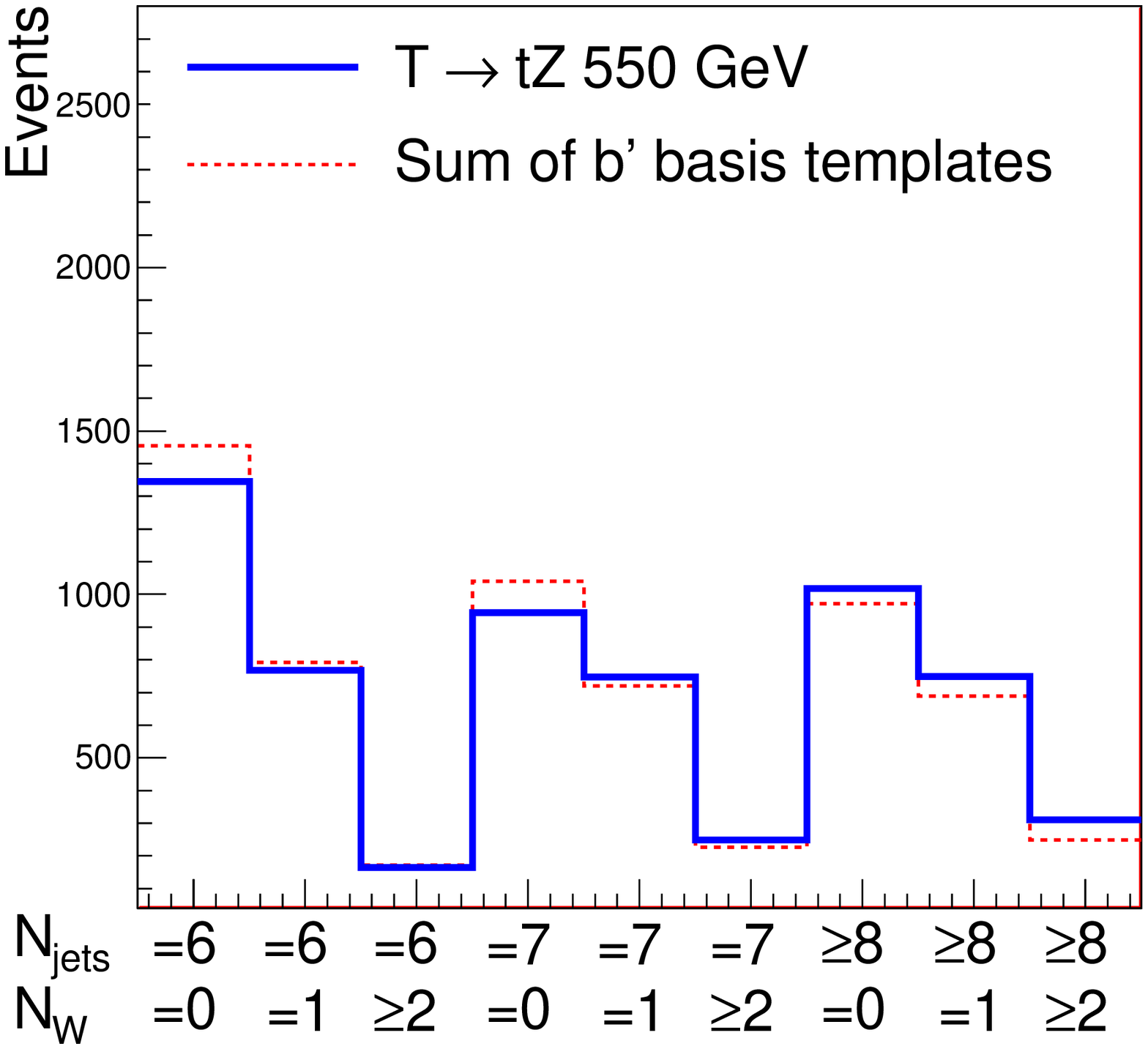}\\
\caption{ Jet and $W$-boson multiplicity, for $T\rightarrow tZ$
  events (solid blue). Also shown (dashed red) is the sum of $b'$ basis templates
  used to derive the limit on the $T$ quark model. Left for $m_T=450$
  GeV, right for $m_T=550$ GeV.}
\label{fig:tZ}
\end{figure}

\begin{figure}[Ht]
\includegraphics[width=0.7\linewidth]{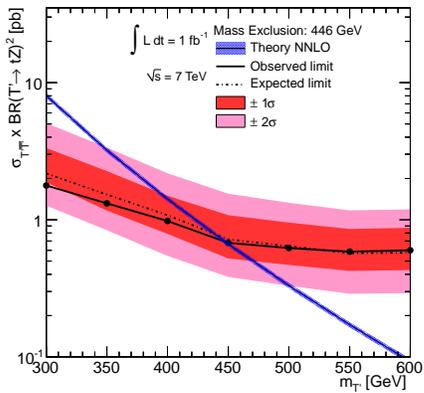}\\
\caption{ Upper limits at 95\% CL on the cross-section for $T$-quark pair
  production in the $tZ$ decay mode.}
\label{fig:lim_tZ}
\end{figure}

The templates for $T\rightarrow tZ$ are shown in Fig.~\ref{fig:tZ} as
well as the linear combination of $b'$ templates used as a set of
basis templates.  Table~\ref{tab:tZ} gives the details of the
decomposition and Fig.~\ref{fig:lim_tZ} shows the derived upper limits
on the cross-section. If BR($T\rightarrow tZ$)=100\%, the
reinterpretation of the $b'$ search yields $m_T>446$ GeV at 95\% CL, comparable
to the CMS limit of $m_T>475$ GeV derived from a search optimized for
this mode.

\begin{table} 
 \caption{ Details of the predicted limit on $T$ pair-production and 
  $T \rightarrow tZ$ decay,  using basis templates from $b'\rightarrow 
  tW$ decays at ATLAS.} 
 \label{tab:tZ}
\begin{tabular}{lrrrrrrr}
\hline\hline 
 &$T_{300}$ & $T_{350}$ & $T_{400}$ & $T_{450}$ & $T_{500}$ & $T_{550}$ & $T_{600}$ \\ \hline

 $a_{b' 300}$  & 0.74  & 0.13  & 0  & 0  & 0  & 0  & 0  \\
 $a_{b' 350}$  & 0  & 0.30  & 0.07  & 0  & 0  & 0  & 0  \\
 $a_{b' 400}$  & 0  & 0.25  & 0.49  & 0.03  & 0  & 0  & 0  \\
 $a_{b' 450}$  & 0  & 0.01  & 0.13  & 0.37  & 0.10  & 0  & 0  \\
 $a_{b' 500}$  & 0  & 0  & 0 & 0.14  & 0.024  & 0.01  & 0  \\
 $a_{b' 550}$  & 0  & 0  & 0  & 0.28  & 0.36  & 0.06  & 0.01  \\
 $a_{b' 600}$  & 0  & 0  & 0  & 0.76  & 0.97  & 1.14  & 0.66  \\\hline 
$\sigma^{\rm limit}_i/\sigma_i^{\rm theory}$ (pred) & 0.22  & 0.41  & 0.69  & 1.02  & 1.87  & 3.4  & 6.47  \\
$\sigma^{\rm limit}$ [pb] (pred) & 1.77  & 1.31  & 0.98  & 0.67  & 0.62  & 0.58  & 0.6  \\ 
 \hline\hline 
 \end{tabular} 
 \end{table}

\subsection{Decays to $th$}

Decays of the $T$-quark to $th$ would give a $t\bar{t}hh$ final state.
If $m_H=125$ GeV, the predominant Higgs boson decay mode is
$b\bar{b}$. As in the case of $tZ$ decays, this gives a final state
similar to that used in the $b'$ search, though the larger Higgs mass
gives a smaller number of observed $W$-boson tags due to the $70-100$ GeV
mass window, and a somewhat larger jet multiplicity.

\begin{figure}[Ht]
\includegraphics[width=0.44\linewidth]{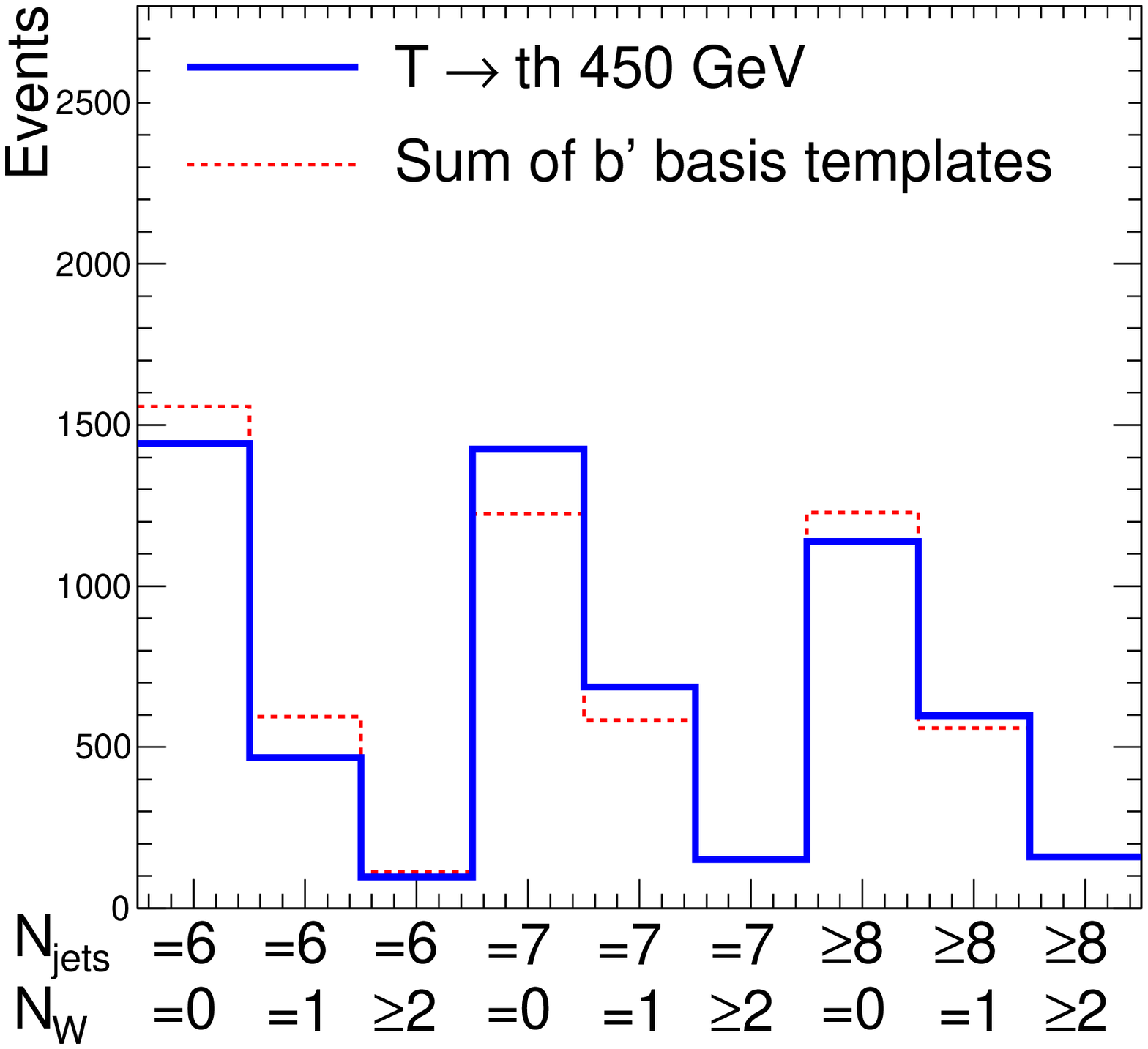}
\includegraphics[width=0.44\linewidth]{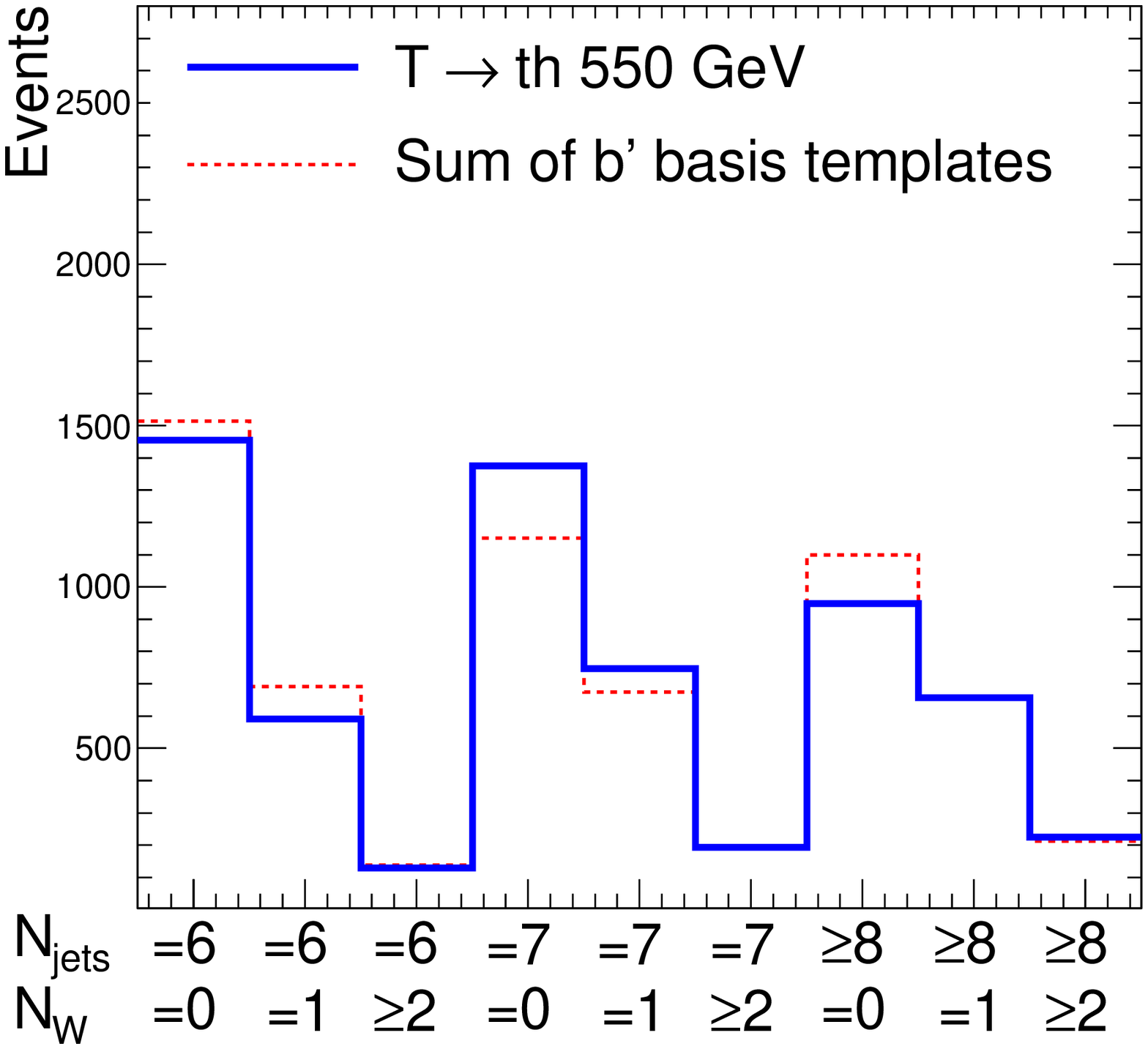}\\
\caption{ Jet and $W$-boson multiplicity, for $T\rightarrow th$
  events (solid blue). Also shown (dashed red) is the sum of $b'$ basis templates
  used to derive the limit on the $T$ quark model. Left for $m_T=450$
  GeV, right for $m_T=550$ GeV.}
\label{fig:tH}
\end{figure}

The templates for $T\rightarrow th$ decay are shown in
Fig.~\ref{fig:tH}. It was not possible to find a linear combination of
$b'$ basis templates which accurately reproduce the $th$ templates
unless we remove $th$ events with more than eight jets. This reduces
the sensitivity, as the overall yield is decreased, and thus produces a somewhat
conservative limit.

\begin{figure}[Ht]
\includegraphics[width=0.7\linewidth]{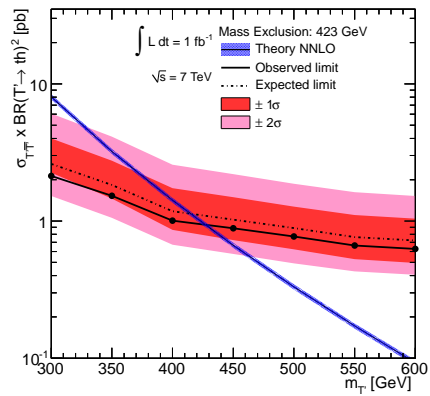}\\
\caption{  Upper limits at 95\% CL on the cross-section for $T$-quark pair  production in the $th$ decay mode.}
\label{fig:lim_tH}
\end{figure}

 Table~\ref{tab:tH} gives the details of the
decomposition and Fig.~\ref{fig:lim_tH} shows the derived upper limits
on the cross-section. If BR($T\rightarrow th$)=100\%, the
reinterpretation of the $b'$ search yields $m_T>423$ GeV at 95\% CL, the first
limit in this mode.

\begin{table} 
 \caption{ Details of the predicted limit on $T$ pair-production and 
  $T\rightarrow tH$ decay,  using basis templates from $b'\rightarrow 
  Wt$ decays at ATLAS.} 
 \label{tab:tH}
\begin{tabular}{lrrrrrrr}
\hline\hline 
 &$T_{300}$ & $T_{350}$ & $T_{400}$ & $T_{450}$ & $T_{500}$ & $T_{550}$ & $T_{600}$ \\ \hline

 $a_{b' 300}$  & 0.62  & 0.23  & 0.07  & 0.02  & 0  & 0  & 0  \\
 $a_{b' 350}$  & 0  & 0  & 0  & 0.03  & 0  & 0  & 0  \\
 $a_{b' 400}$  & 0  & 0  & 0  & 0.01  & 0.03  & 0  & 0  \\
 $a_{b' 450}$  & 0  & 0  & 0  & 0  & 0  & 0  & 0  \\
 $a_{b' 500}$  & 0  & 0.77  & 1.13  & 0.62  & 0.40  & 0.29  & 0.17  \\
 $a_{b' 550}$  & 0  & 0  & 0  & 0  & 0  & 0  & 0  \\
 $a_{b' 600}$  & 0  & 0.03  & 0  & 0  & 0  & 0.01  & 0  \\\hline 
$\sigma^{\rm limit}_i/\sigma_i^{\rm theory}$ (pred) & 0.26  & 0.47  & 0.7  & 1.32  & 2.32  & 3.85  & 6.75  \\
$\sigma^{\rm limit}$ [pb] (pred) & 2.13  & 1.52  & 1  & 0.88  & 0.77  & 0.66  & 0.62  \\ 
 \hline\hline 
 \end{tabular} 
 \end{table}

\subsection{Decays to $tZ$ and $th$}

To probe a mixed case, we allow both $tZ$ and $th$ decays, which gives
 $t\bar{t}ZZ$, $t\bar{t}hh$ modes as well as the mixed mode
 $t\bar{t}Zh$. The relative $tZ:th$ branching ratios are unchanged,
 see Fig.~\ref{fig:br}.

\begin{figure}[Ht]
\includegraphics[width=0.44\linewidth]{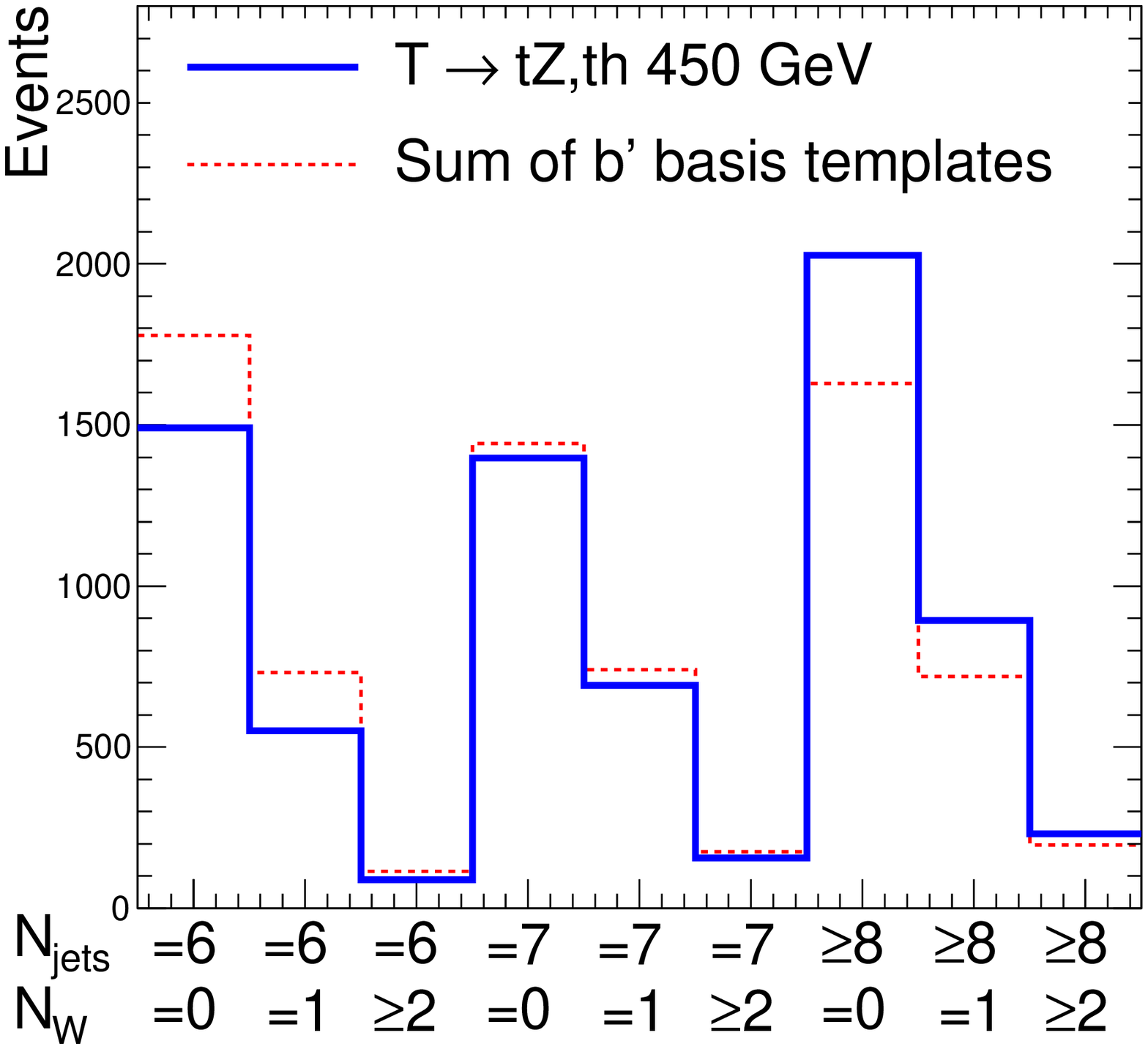}
\includegraphics[width=0.44\linewidth]{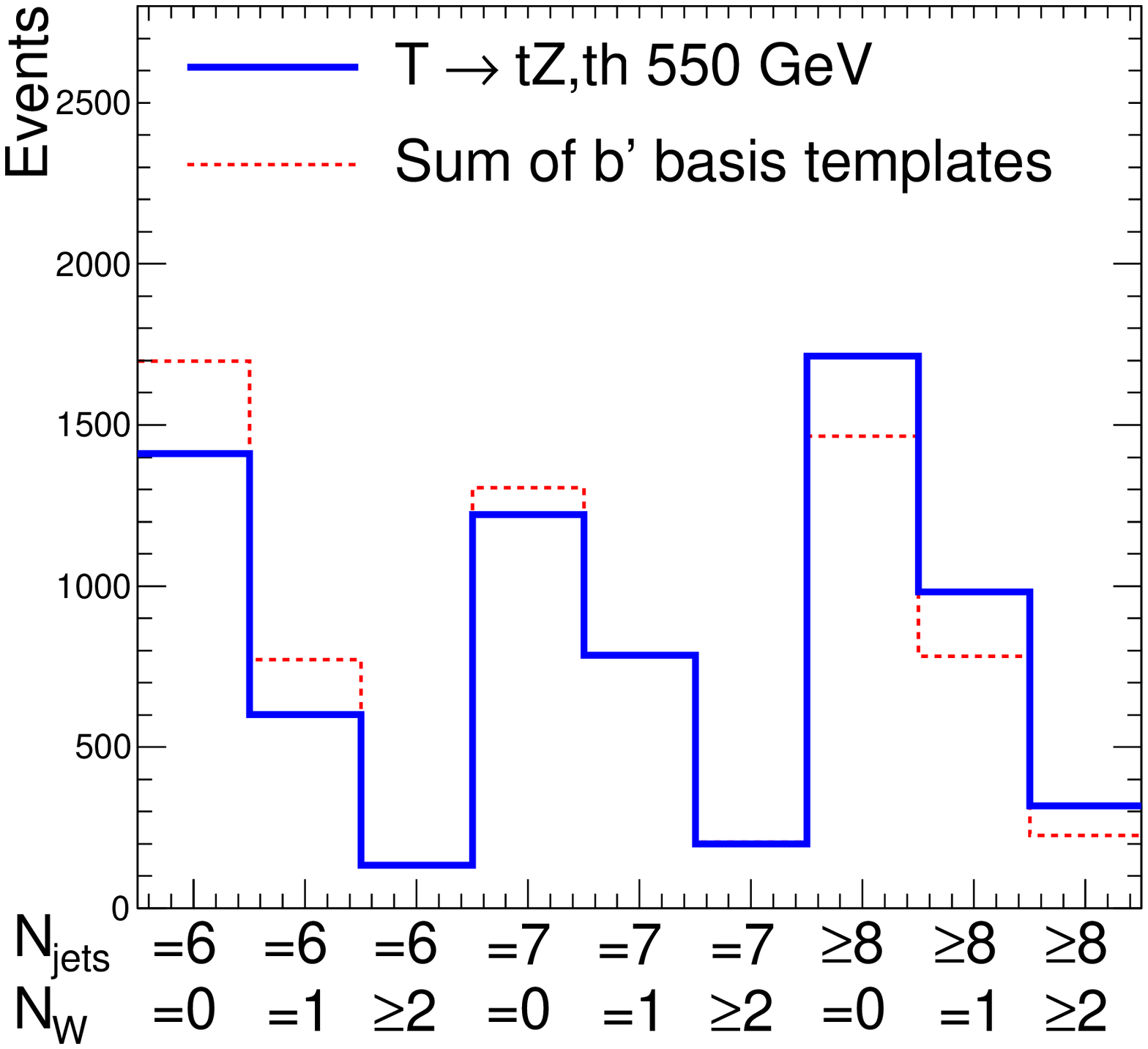}\\
\caption{ Jet and $W$-boson multiplicity, for $T\rightarrow tZ,th$
  events (solid blue). Also shown (dashed red) is the sum of $b'$ basis templates
  used to derive the limit on the $T$ quark model. Left for $m_T=450$
  GeV, right for $m_T=550$ GeV.}
\label{fig:tZtH}
\end{figure}

\begin{figure}[Ht]
\includegraphics[width=0.7\linewidth]{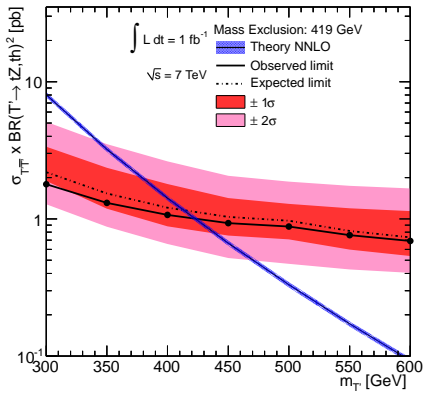}\\
\caption{  Upper limits at 95\% CL on the cross-section for $T$-quark pair
  production in the $tZ, tH$ decay mode.}
\label{fig:lim_tZtH}
\end{figure}

The templates for $T\rightarrow th,tZ$ decay are shown in
Fig.~\ref{fig:tZtH}. 
 Table~\ref{tab:tZtH} gives the details of the
decomposition and Fig.~\ref{fig:lim_tZtH} shows the derived upper limits
on the cross-section. If BR($th$ or $tZ$)=100\%, the
reinterpretation of the $b'$ search yields $m_T>419$ GeV at 95\% CL, the first
limit in this mixed mode.

\begin{table} 
 \caption{ Details of the predicted limit on $T$ pair-production and 
  $T\rightarrow tZ,tH$ decays,  using basis templates from $b'\rightarrow 
  Wt$ decays at ATLAS.} 
 \label{tab:tZtH}
\begin{tabular}{lrrrrrrr}
\hline\hline 
 &$T_{300}$ & $T_{350}$ & $T_{400}$ & $T_{450}$ & $T_{500}$ & $T_{550}$ & $T_{600}$ \\ \hline

 $a_{b' 300}$  & 0.74  & 0.15  & 0  & 0  & 0  & 0  & 0  \\
 $a_{b' 350}$  & 0  & 0.41  & 0.27  & 0.05  & 0  & 0  & 0  \\
 $a_{b' 400}$  & 0  & 0  & 0.15  & 0.26  & 0.18  & 0.05  & 0.00  \\
 $a_{b' 450}$  & 0  & 0  & 0  & 0  & 0.01  & 0.10  & 0.09  \\
 $a_{b' 500}$  & 0  & 0  & 0  & 0  & 0  & 0  & 0.00  \\
 $a_{b' 550}$  & 0  & 0  & 0  & 0  & 0  & 0  & 0  \\
 $a_{b' 600}$  & 0  & 0  & 0  & 0  & 0  & 0  & 0  \\\hline 
$\sigma^{\rm limit}_i/\sigma_i^{\rm theory}$ (pred) & 0.22  & 0.4  & 0.75  & 1.4  & 2.65  & 4.42  & 7.45  \\
$\sigma^{\rm limit}$ [pb] (pred) & 1.79  & 1.3  & 1.06  & 0.93  & 0.88  & 0.76  & 0.69  \\ 
 \hline\hline 
 \end{tabular} 
 \end{table}

\subsection{Decays with $Wb$}

\begin{figure}[Ht]
\includegraphics[width=0.44\linewidth]{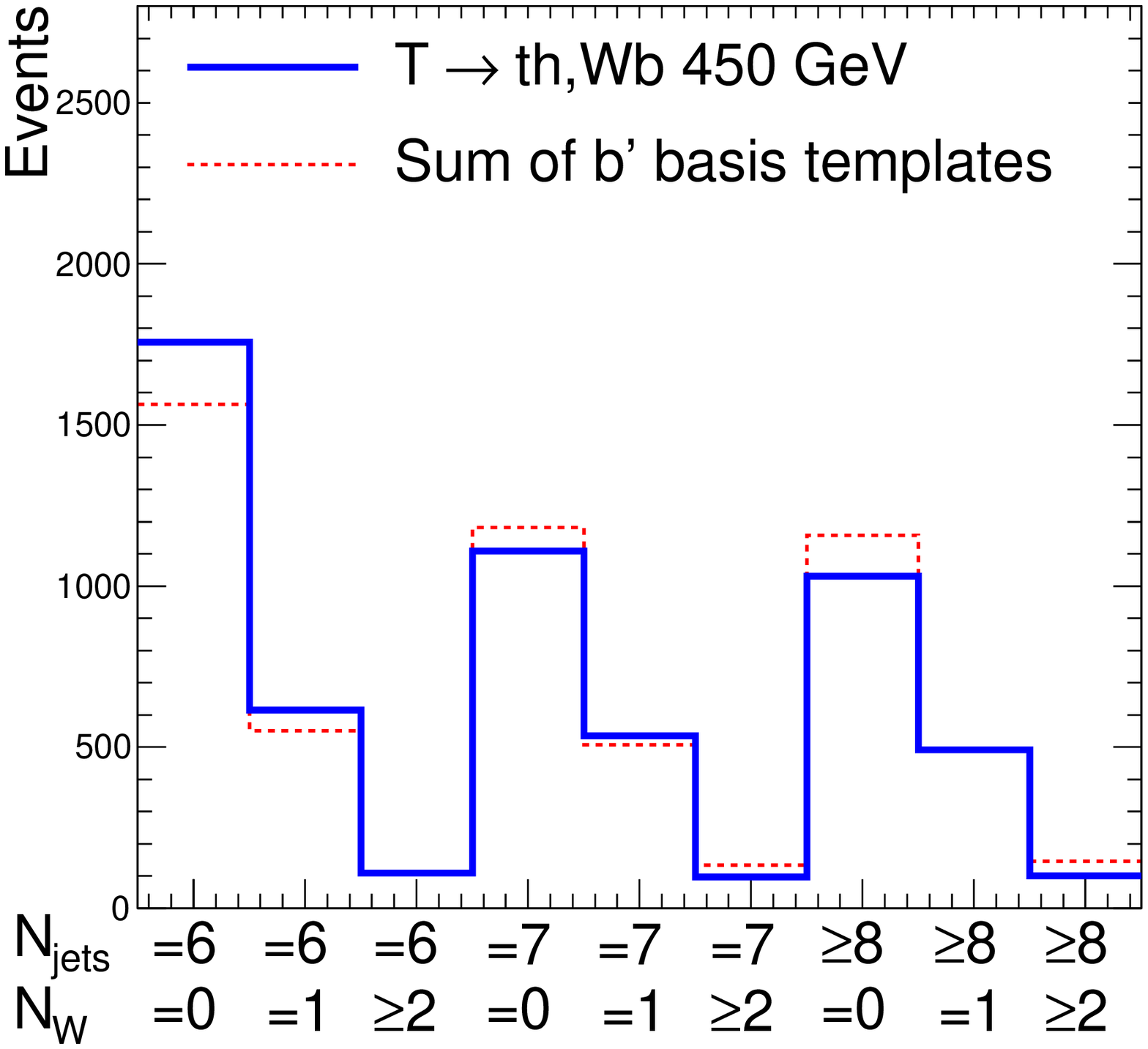}
\includegraphics[width=0.44\linewidth]{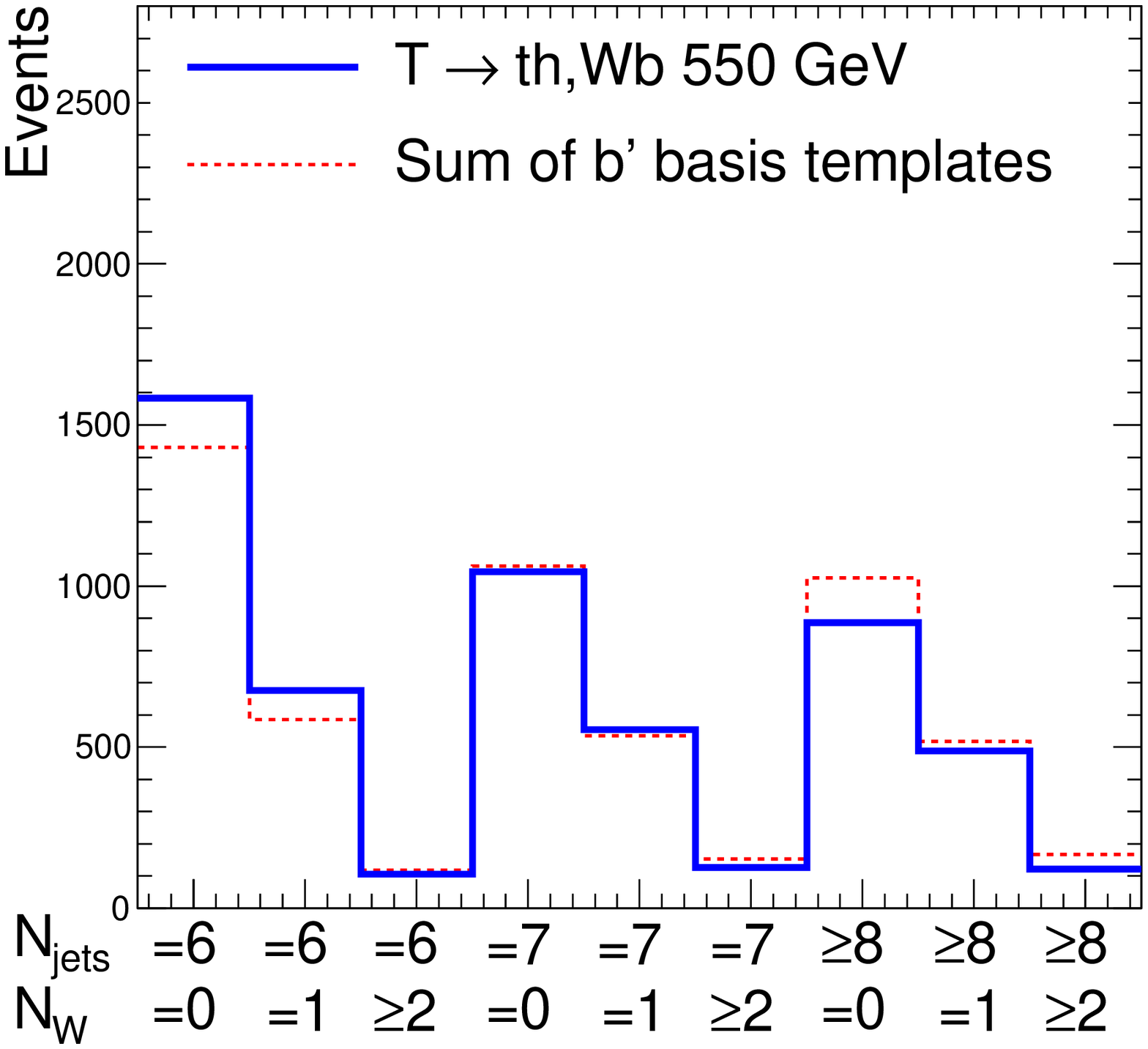}\\
\caption{ Jet and $W$-boson multiplicity, for $T\rightarrow Wb,th$
  events (solid blue). Also shown (dashed red) is the sum of $b'$ basis templates
  used to derive the limit on the $T$ quark model. Left for $m_T=450$
  GeV, right for $m_T=550$ GeV.}
\label{fig:tHWb}
\end{figure}

\begin{figure}[Ht]
\includegraphics[width=0.7\linewidth]{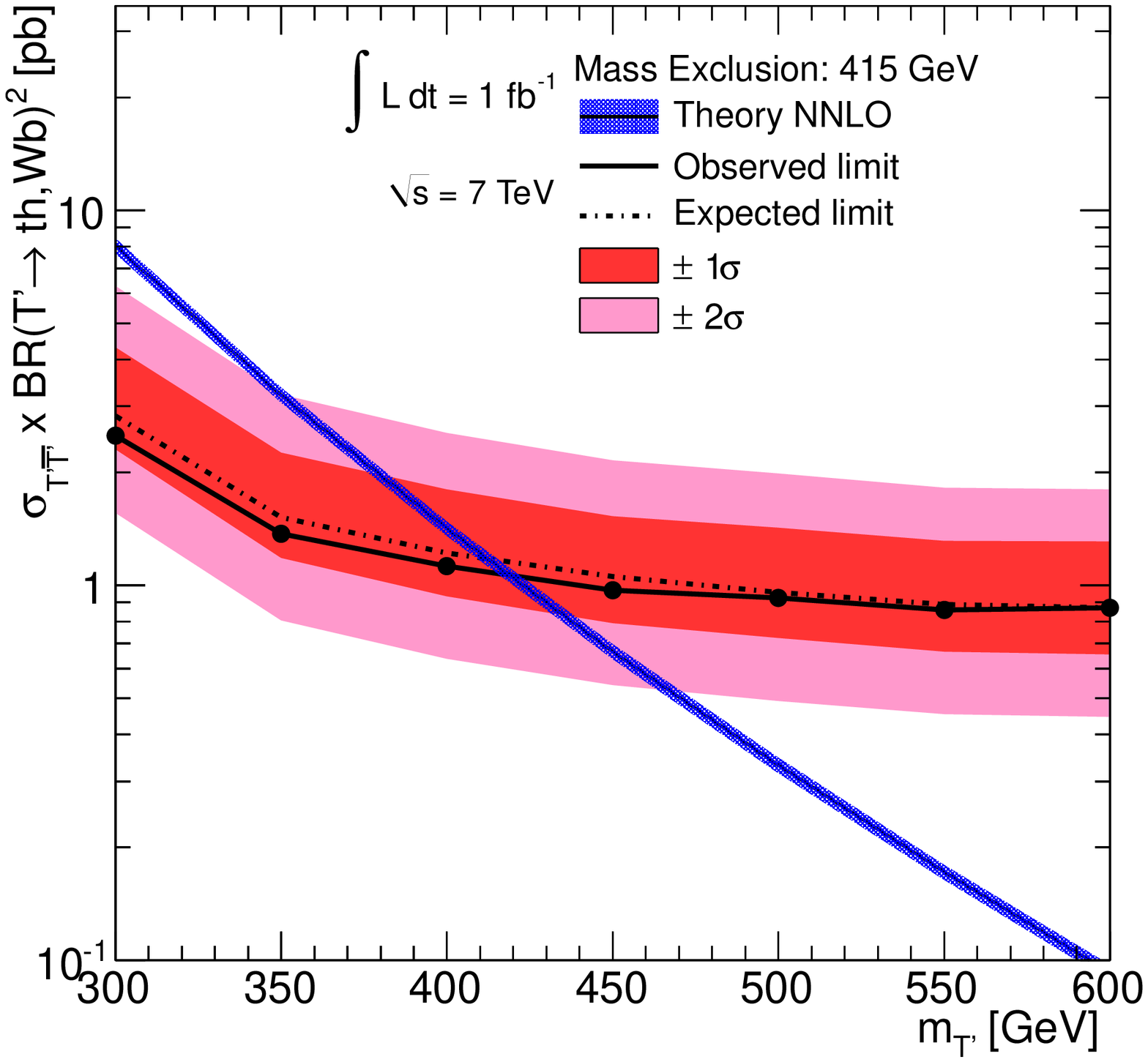}\\
\caption{ 
 Upper limits at 95\% CL on the cross-section for $T$-quark  pair  production in the $th, Wb$ decay mode.}
\label{fig:lim_tHWb}
\end{figure}

We probe two more mixed-decay cases which include $Wb$ decays. The
first allows either $Wb$ or $th$ decays; see Fig.~\ref{fig:tHWb} for
templates, Table~\ref{tab:tHWb} for decomposition details and
Fig.~\ref{fig:lim_tHWb} for limits.   If BR($th$ or $Wb$)=100\%, the
reinterpretation of the $b'$ search yields $m_T>415$ GeV at 95\% CL, the first
limit in this mixed mode.  

\begin{figure}[Ht]
\includegraphics[width=0.44\linewidth]{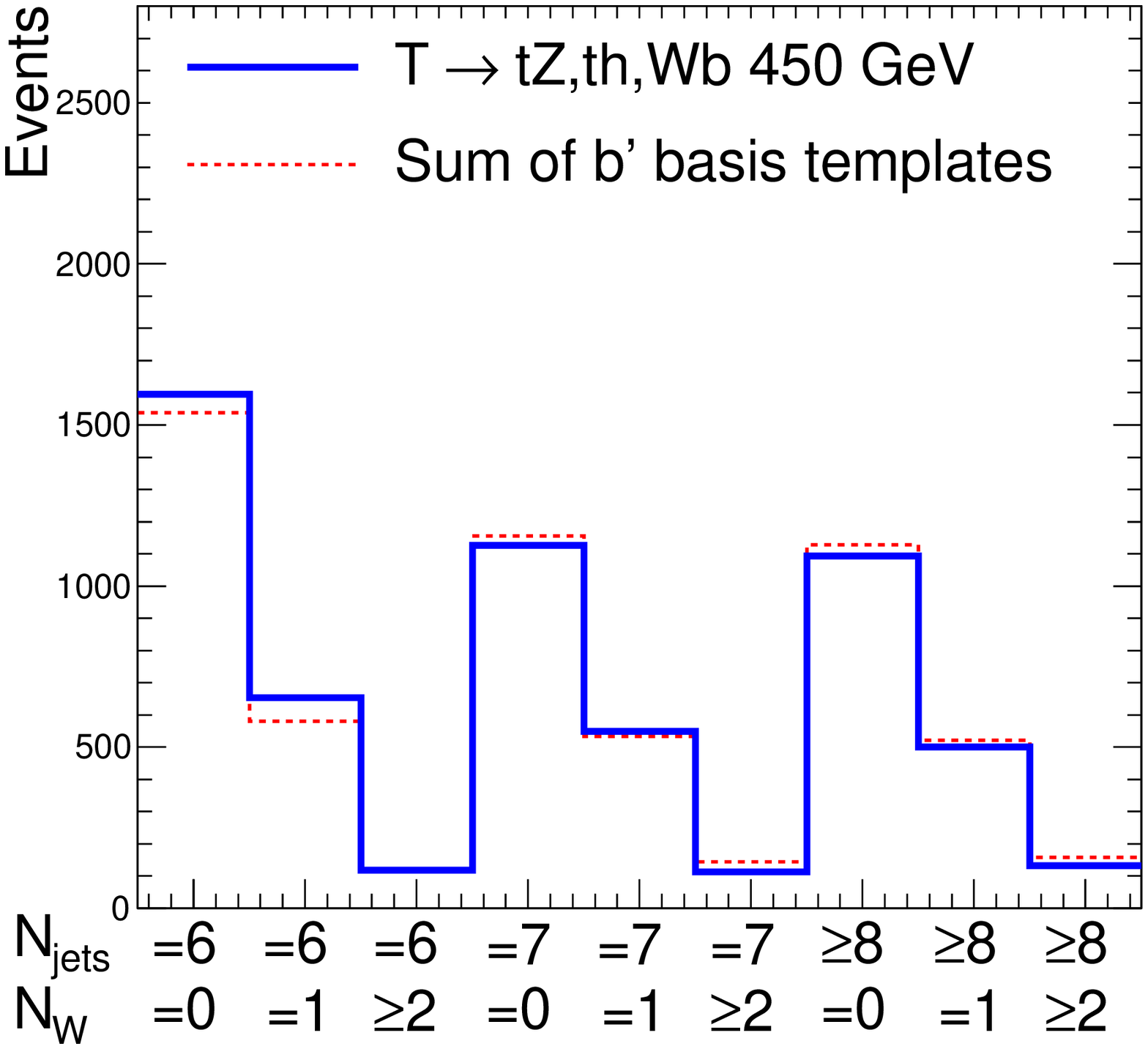}
\includegraphics[width=0.44\linewidth]{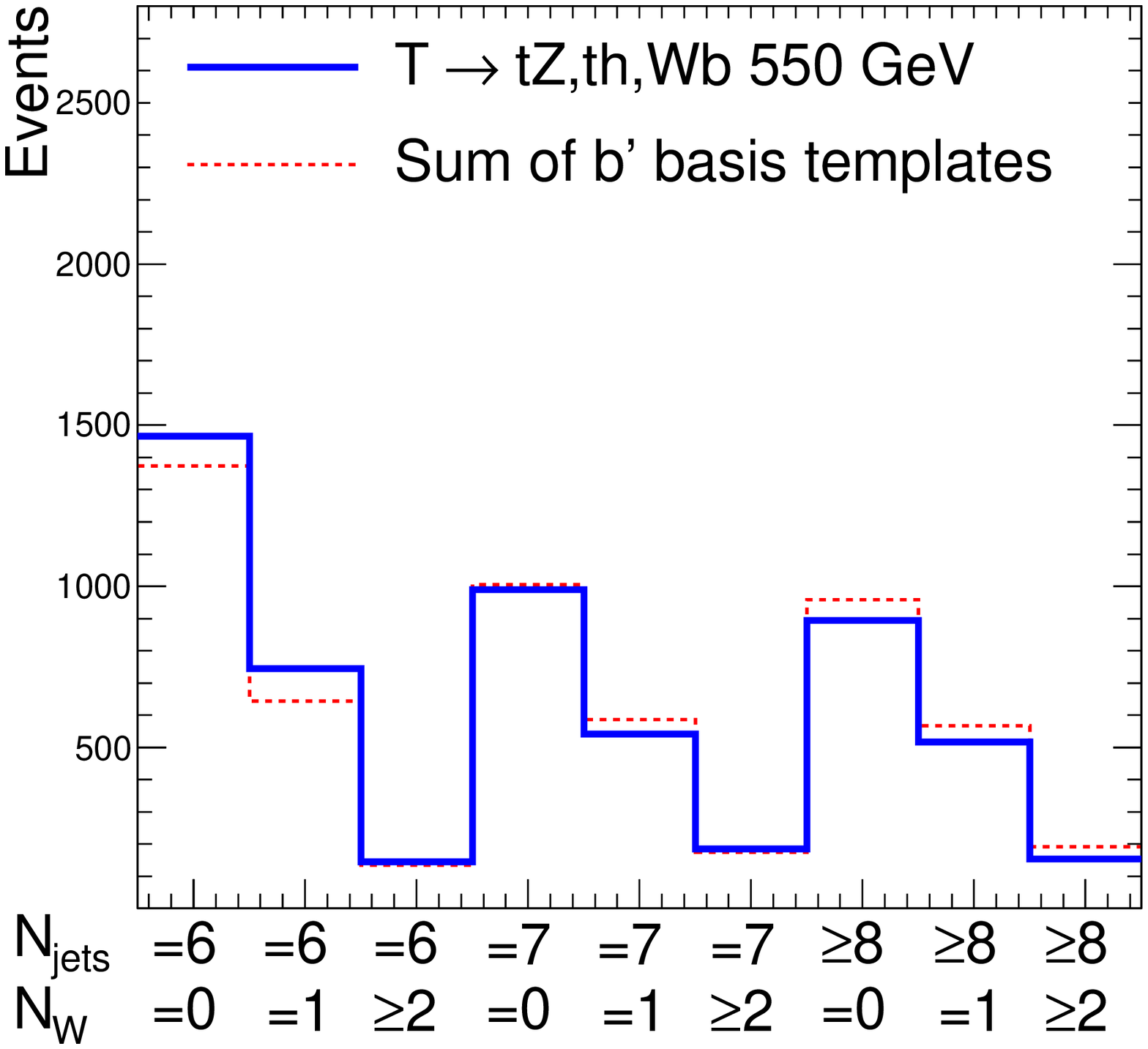}\\
\caption{ Jet and $W$-boson multiplicity, for $T\rightarrow Wb,th,tZ$
  events (solid blue). Also shown (dashed red) is the sum of $b'$ basis templates
  used to derive the limit on the $T$ quark model. Left for $m_T=450$
  GeV, right for $m_T=550$ GeV.}
\label{fig:all}
\end{figure}

\begin{figure}[Ht]
\includegraphics[width=0.7\linewidth]{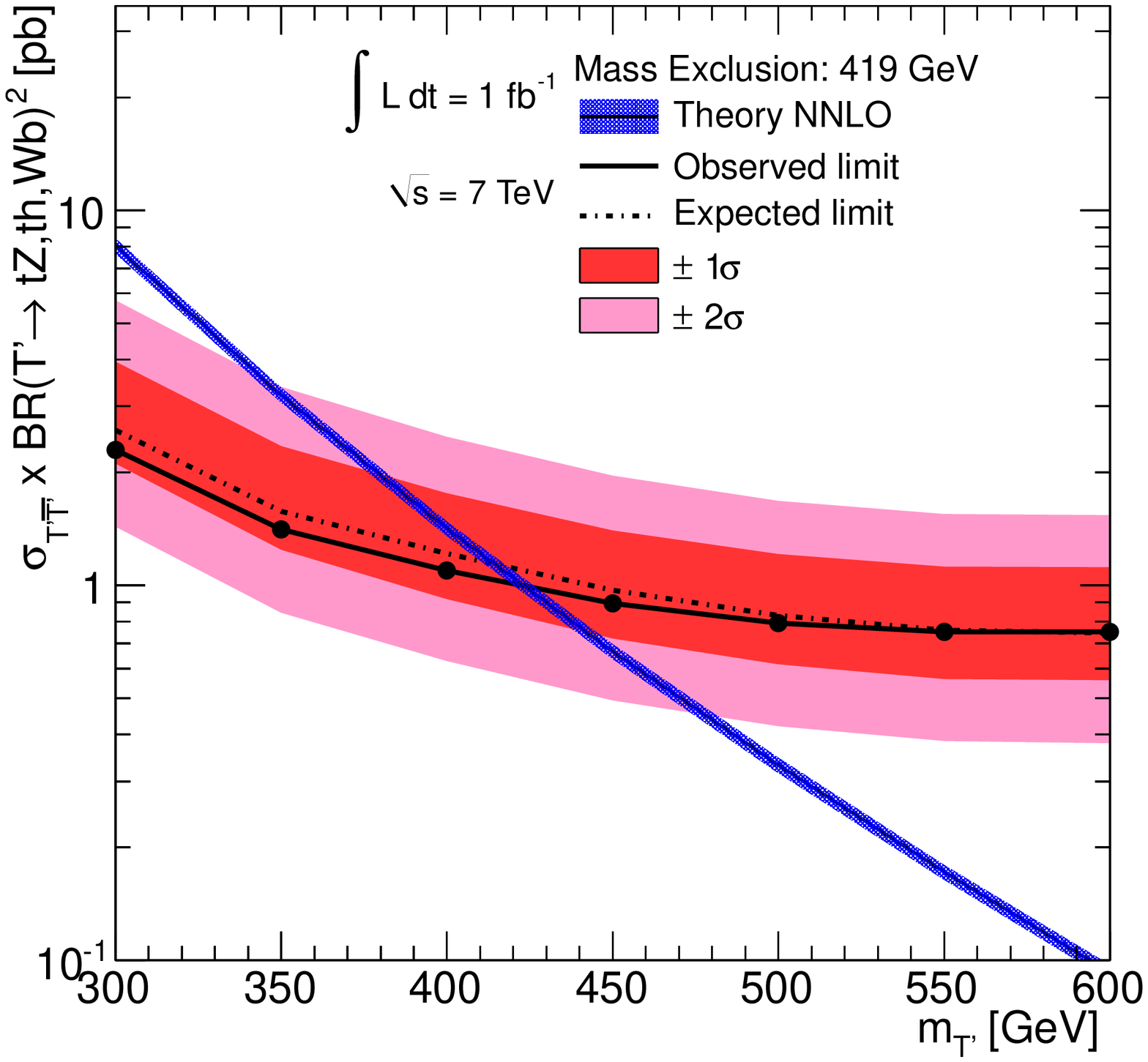}\\
\caption{  Upper limits at 95\% CL on the cross-section for $T$-quark
  pair  production in the $th, tZ, Wb$ decay mode.}
\label{fig:lim_all}
\end{figure}

The second case allows all
three decays ($Wb$, $th$ or $tZ$) in the predicted mixture
(Fig.~\ref{fig:br}); see Fig.~\ref{fig:all} for
templates, Table~\ref{tab:all} for decomposition details and
Fig.~\ref{fig:lim_all} for limits. If BR($th$ or $tZ$ or $Wb$)=100\%, the
reinterpretation of the $b'$ search yields $m_T>419$ GeV at 95\% CL as reported previously~\cite{basis}.

In both cases, the $T$-quark signal has a jet multiplicity which is
quite different from the $b'\rightarrow Wt$ basis templates.  The most
accurate description of $T$-quark templates uses the low-mass and
high-mass $b'$ basis templates
(see Tables~\ref{tab:tHWb} and~\ref{tab:all})
which have the least overlap and so can be varied nearly independently to
achieve the desired jet multiplicity of the $T$-quark signals with
$Wb$ decays.  

\begin{table} 
 \caption{ Details of the predicted limit on $T$ pair-production and 
  $T\rightarrow  tH,Wb$ decays,  using basis templates from $b'\rightarrow Wt$ decays at ATLAS.} 
 \label{tab:tHWb}
\begin{tabular}{lrrrrrrr}
\hline\hline 
 &$T_{300}$ & $T_{350}$ & $T_{400}$ & $T_{450}$ & $T_{500}$ & $T_{550}$ & $T_{600}$ \\ \hline

 $a_{b' 300}$  & 0.36  & 0.18  & 0.07  & 0.03  & 0.01  & 0  & 0  \\
 $a_{b' 350}$  & 0  & 0  & 0  & 0  & 0  & 0  & 0  \\
 $a_{b' 400}$  & 0  & 0  & 0  & 0  & 0  & 0  & 0  \\
 $a_{b' 450}$  & 0  & 0  & 0  & 0  & 0  & 0  & 0  \\
 $a_{b' 500}$  & 0  & 0  & 0  & 0  & 0  & 0  & 0  \\
 $a_{b' 550}$  & 0.17  & 1.18  & 0.72  & 0.68  & 0.15  & 0.13  & 0.03  \\
 $a_{b' 600}$  & 4.07  & 3.27  & 2.16  & 0.97  & 0.95  & 0.51  & 0.35  \\\hline 
$\sigma^{\rm limit}_i/\sigma_i^{\rm theory}$ (pred) & 0.31  & 0.42  & 0.79  & 1.45  & 2.79  & 5  & 9.37  \\
$\sigma^{\rm limit}$ [pb] (pred) & 2.5  & 1.37  & 1.12  & 0.96  & 0.92  & 0.86  & 0.86  \\ 
 \hline\hline 
 \end{tabular} 
 \end{table}

\begin{table} 
 \caption{ Details of the predicted limit on $T$ pair-production and 
  $tZ,tH,Wb$ decays,  using basis templates from $b'\rightarrow 
  tW$ decays at ATLAS.} 
 \label{tab:all}
\begin{tabular}{lrrrrrrr}
\hline\hline 
 &$T_{300}$ & $T_{350}$ & $T_{400}$ & $T_{450}$ & $T_{500}$ & $T_{550}$ & $T_{600}$ \\ \hline

 $a_{b' 300}$  & 0.39  & 0.20  & 0.07  & 0.02  & 0.01  & 0  & 0  \\
 $a_{b' 350}$  & 0  & 0  & 0  & 0  & 0  & 0  & 0  \\
 $a_{b' 400}$  & 0  & 0  & 0  & 0  & 0  & 0  & 0  \\
 $a_{b' 450}$  & 0  & 0  & 0  & 0  & 0  & 0  & 0  \\
 $a_{b' 500}$  & 0.01  & 0  & 0  & 0  & 0  & 0  & 0  \\
 $a_{b' 550}$  & 0.24  & 0.94  & 1.24  & 0.78  & 0.39  & 0.13  & 0.04  \\
 $a_{b' 600}$  & 4.35  & 2.88  & 1.46  & 1.11  & 0.87  & 0.68  & 0.42  \\\hline 
$\sigma^{\rm limit}_i/\sigma_i^{\rm theory}$ (pred) & 0.28  & 0.43  & 0.77  & 1.34  & 2.38  & 4.36  & 8.1  \\
$\sigma^{\rm limit}$ [pb] (pred) & 2.29  & 1.4  & 1.09  & 0.89  & 0.79  & 0.75  & 0.75  \\ 
 \hline\hline 
 \end{tabular} 
 \end{table}

\begin{figure}[!hb]
\includegraphics[width=0.7\linewidth]{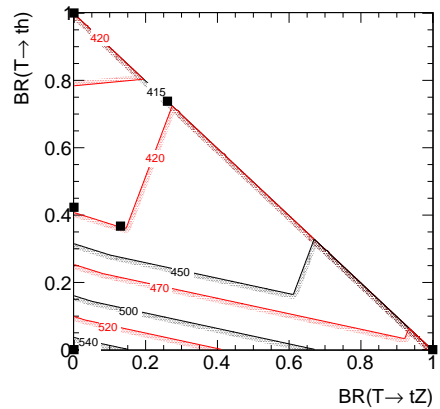}\\
\caption{ Lower limits at 95\% CL on the mass (GeV) of the exotic $T$ quark
  for varying  assumptions about branching ratios.  Black squares show
  the points at which the limits are calculated; contours are interpolated. Limits at
  BR($Wb$)=100\%~\cite{cmstpll} and BR($tZ$)=100\%~\cite{cmstz} are taken from dedicated searches;
other points our reinterpretations (see
  Table~\ref{tab:sum}). }
\label{fig:lim_tri}
\end{figure}

\section{Limits in the branching ratio triangle}

Each of the results above place lower limits on the mass of the $T$
quark under specific assumptions about the branching ratios.
Table~\ref{tab:sum} summarizes these results. For each scenario, we quote the branching ratios at the limit value.
Figure~\ref{fig:lim_tri} shows the results graphically, with
interpolated values.

\begin{table}
\caption{Summary of mass limits at 95\% CL for various branching ratios. For
  BR(tZ)=100\%, we include for comparison the direct limit from CMS~\cite{cmstz} as
  well as our reinterpretation of the ATLAS~\cite{bprime}
  $b'\rightarrow Wt$ search for this case. Uncertainties are due to
  imperfect description of the signal model via the basis templates. }
\label{tab:sum}
\begin{tabular}{lllrl}
\hline\hline 
& & & Lower limit & \\
BR($Wb$) &BR($tZ$) &BR($th$) &
$m_T$ [GeV] & Comment \\
\hline
\vspace{1mm} 1 & 0 & 0 & 557 & Direct search~\cite{cmstpll} \\
\vspace{1mm} 0 & 1 & 0 & 475 & Direct search~\cite{cmstz} \\
\vspace{1mm} 0 & 1 & 0 & $446^{+4}_{-4}$ & Reinterpreted here\\
\vspace{1mm} 0 & 0 & 1 & $423^{+23}_{-48}$ & Reinterpreted here\\
\vspace{1mm} 0.50 & 0.13 & 0.37 & $419^{+3}_{-3}$ & Reinterpreted~\cite{basis} \\
\vspace{1mm} 0 & 0.26 & 0.74 & $419^{+12}_{-18}$ & Reinterpreted here \\
\vspace{1mm} 0.58 & 0 & 0.42 & $415^{+6}_{-11}$ & Reinterpreted here\\
\hline\hline 
\end{tabular}
\end{table}

\section{Conclusions}

Searches for an exotic heavy quark $T$ have been previously reported for
specific choices of the decay modes: BR($Wb$)=100\% or
BR($tZ$)=100\%.  We consider alternative branching ratio scenarios,
scanning the branching ratio triangle. We derive limits for these
scenarios by reinterpretting a recent ATLAS search for $b'$.

We find limits for $T$ at $m_T>415$ GeV across the entire triangle, up
to $m_T>557$ GeV in the case of $Wb$ decay.  An optimized experimental
search for decays with large $th$ branching ratio would lead to
significantly tighter limits across the triangle.

\section{Acknowledgements}

We thank Michael Peskin, Eric Albin, Michael Werth and Nadine Amsel for useful comments.
We thank  Graham Kribs and Adam Martin for the $T$ quark model and technical support.
The authors are supported by grants from the Department of Energy
Office of Science and by the Alfred P. Sloan Foundation.


\clearpage

\begin{thebibliography}{99}

\bibitem{cmstpll} 
CMS Collaboration,
  %``Search for heavy, top-like quark pair production in the dilepton final state in pp collisions at sqrt(s) = 7 TeV,''
  arXiv:1203.5410 (2011).

\bibitem{cmsbpll} CMS Collaboration, arXiv:1204.1088 (2012).

\bibitem{flacco}
C.~J.~Flacco, D.~Whiteson, T.~M.~P.~Tait and S.~Bar-Shalom,
  %``Direct Mass Limits for Chiral Fourth-Generation Quarks in All Mixing Scenarios,''
  Phys.\ Rev.\ Lett.\  {\bf 105}, 111801 (2010)
  [arXiv:1005.1077 [hep-ph]].

\bibitem{tprime}   G.~D.~Kribs, A.~Martin and T.~S.~Roy,
  %``Higgs boson discovery through top-partners decays using jet substructure,''
  Phys.\ Rev.\ D {\bf 84}, 095024 (2011)
  [arXiv:1012.2866 [hep-ph]].

\bibitem{jaas}
  J.~A.~Aguilar-Saavedra,
  %``Identifying top partners at LHC,''
  JHEP {\bf 0911}, 030 (2009).

\bibitem{thiggs} M. Perelstein, M. Peskin, A. Pierce, Phys.Rev. D69 (2004) 075002.

\bibitem{littlehiggs} N. Arkani-Hamed  A.G. Cohen, E. Katz,  A.E. Nelson, JHEP 0207 (2002) 034.

\bibitem{littlehiggs2} N. Arkani-Hamed, A.G. Cohen , E. Katz,
  A.E. Nelson, T. Gregoire, Jay G. Wacker, JHEP 0208 (2002) 021

\bibitem{littlehiggs3} T. Han, H. Logan, B. McElrath, L.-T. Wang, Phys.Rev. D67 (2003) 095004.



\bibitem{atlasstudy} G. Azuelos {\it et. al.} Eur.Phys.J. C39S2 (2005).


\bibitem{cmstz}
CMS Collaboration,
  %``Search for a Vector-like Quark with Charge 2/3 in t + Z Events from pp Collisions at sqrt(s) = 7 TeV,''
  Phys.\ Rev.\ Lett.\  {\bf 107}, 271802 (2011)
  [arXiv:1109.4985 [hep-ex]].



\bibitem{basis}
 K.~Rao and D.~Whiteson,
  arXiv:1203.6642 [hep-ex] (2012).

\bibitem{atlastplj} ATLAS Collaboration, arXiv:1202.3076 (2012).

\bibitem{atlastpll} ATLAS Collaboration, arXiv:1202.3389 (2012).

\bibitem{bat} T. Nagel, Phil. Rev., 435 (1974).

\bibitem{bprime} ATLAS Collaboration,  arXiv:1202.6540 (2012).


\bibitem{pgs} J. Conway,  {\tiny
    \mbox{\texttt{http://www.physics.ucdavis.edu/\~ conway/research/software/pgs/pgs.html}}}.

\bibitem{atlasbp} ATLAS Collaboration, arXiv:1202.6540 (2012).

\bibitem{cls1} {A. Read},   J. Phys. G: Nucl. Part. Phys. {\bf 28}, 2693 (2002);
\bibitem{cls2} {T. Junk},  Nucl. Instrum. Methods A {\bf 434}, 425
  (1999).

\bibitem{geant} S. Agostinelli {\it et al.}, Nucl. Inst.  Meth. {\bf A}506 (2003) 250-303.



\bibitem{madgraph}
  J.~Alwall, M.~Herquet, F.~Maltoni, O.~Mattelaer and T.~Stelzer,
  %``MadGraph 5 : Going Beyond,''
  JHEP {\bf 1106}, 128 (2011)
  [arXiv:1106.0522 [hep-ph]].
  %%CITATION = ARXIV:1106.0522;%%

\bibitem{pythia}
  T.~Sjostrand, S.~Mrenna and P.~Z.~Skands,
  %``PYTHIA 6.4 Physics and Manual,''
  JHEP {\bf 0605}, 026 (2006)
  [hep-ph/0603175].
  %%CITATION = HEP-PH/0603175;%%

\bibitem{d4xs} M. Aliev {\it et al.}, Comput. Phys. Commun. {\bf 182}, 1034 (2011).



\end{thebibliography}
\end{document}